\documentclass{nature}
\usepackage{graphicx}
\usepackage{amssymb}
\usepackage{amsmath}

\title{Fundamental limits to nonlinear energy harvesting}

\author{Ashkan Haji Hosseinloo \& Konstantin Turitsyn}

\begin{document}

\maketitle

\begin{affiliations}
 \item[] Department of Mechanical Engineering, Massachusetts Institute of Technology, 77 Massachusetts Avenue, Cambridge, Massachusetts 02139, USA
\end{affiliations}

\begin{abstract}
Ease of miniaturization, and less or no maintenance, among other advantages, have pushed towards replacement of conventional batteries with energy harvesters in particular, vibratory energy harvesters. In the recent years, nonlinearity has been intentionally introduced into the otherwise linear energy harvesters in the hope of increasing the frequency bandwidth and power density.  However, fundamental limits on the harvestable energy of a harvester subjected to an arbitrary excitation force is yet unknown. Understanding of these limits is not only essential for assessment of the technology potential, but also provides a broader prospective on the current harvesting mechanisms and guidance in their improvement. Here we derive the fundamental limits on output power of an ideal energy harvester, and develop an analysis framework for simple computation of this limit for more sophisticated set-ups. We show that the optimal harvester maximizes the harvested energy through a mechanical analogue of ''buy low-sell high'' strategy. Inspired by this strategy we propose a novel concept of latch-assisted harvesting that is shown to harvest energy more efficiently than its linear and bistable counterparts over a wider range of excitation frequencies and amplitudes.
\end{abstract}

Advances in different fields of technology have continuously reduced the power consumption in electronic devices such as wireless sensors, data transmitters, and medical implants. However, the batteries as the most common power source for electronic devices have not improved much in terms of their energy density over the last couple of decades\cite{paradiso2005energy}. Hence, the problem of energy supply has been one of the biggest issues in miniaturizing many electronic devices. For instance, the battery takes up about $\frac{2}{3}$ of the size of a pacemaker\cite{karami2012powering} that consequently hinders their further miniaturization. In addition to scaling issues, the recharge, or replacement and disposal of the batteries is usually cumbersome, costly, and could entail health-related and environmental complexities\cite{daqaq2014role}.

To remedy the above-mentioned issues and for further miniaturization of electronic devices, energy harvesting has been investigated and considered as a scalable counterpart for batteries. Among many other sources, ambient vibration has captured researchers' attention in the last decade for its being universal and widely available; from vibration in bridges\cite{elvin2006feasibility,galchev2011harvesting} and waves\cite{scruggs2009harvesting,wang2010piezoelectric} to human walking motion\cite{donelan2008biomechanical,rome2005generating,mitcheson2008energy} and his internal organs\cite{karami2012powering,dagdeviren2014conformal}. A typical vibratory energy harvester (VEH) consists of a vibrating host structure, a transducer, and an electrical load. A broad variety of different electromagnetic, electrostatic, piezoelectric, and magnetostrictive transduction mechanisms have been exploited in VEHs to convert the vibration energy of the host structure into useful electrical energy\cite{erturk2011broadband}.

Most of the conventional VEHs exploit linear resonance to maximize the harvesting efficiency. The natural frequency of the host structure is tuned to the excitation frequency of the harvester. This approach has three obvious downsides. First, linear resonance is inherently narrow in bandwidth, especially if the structure has low damping; hence they are easily detuned by manufacturing tolerances or small changes in excitation frequency. Second, many if not most of the real-world vibrations are driven by very turbulent focrces that have a very broad range of frequencies dominated by the slow ones \cite{frisch1995turbulence}. As a result, the excitation sources  are typically wideband having their power distributed over a wide range of frequencies and also non-stationary having their dominant frequency peaks changing over time\cite{daqaq2014role}. Finally, most of the vibration energy of large scale host structures is contained in low frequency band of the spectrum. At the same time miniature harvesters usually have high natural frequencies, and tuning linear VEHs to low frequency excitation at small scale is another big issue of the linear harvesters. These consequently render linear VEHs quite ineffective in practice.

To overcome the above-mentioned issues, in  addition to resonance tuning, multimodal energy harvesting, and frequency up-conversion\cite{tang2010toward}, in the last few years researchers have tried to make use of purposeful introduction of nonlinearity in VEH design\cite{daqaq2014role}. Different studies have investigated hardening and softening monostable\cite{tang2012improving} or bistable \cite{cottone2009nonlinear,harne2013review,pellegrini2013bistable} energy harvesters. One of the key challenges in designing nonlinear harvesters is the immense range of possible nonlinearities. Unlike linear systems, the nonlinear dynamics systems can not be characterized by small number of parameters and are much more difficult to analyze. In this article we address this challenge by taking a different approach to the design of harvesting devices. Instead of trying different inherent or purposefully-introduced nonlinearities, we seek to find fundamental limits on harvestable energy for an arbitrary excitation source and maximally general harvester structure.

Explicit identification of fundamental performance limits played a crucial role in many fields of science and engineering. In energy field, the classical Carnot cycle efficiency \cite{carnot2005reflections} was a guiding principle for development of thermal power plants, and combustion engines. At the same it has inspired scientific debates and consequently lead to the formation of modern statistical physics. The Lanchester-Betz limit for wind harvesting efficiency \cite{bergey1979lanchester}, and Shockley-Queisser limit for the efficiency of solar cells\cite{shockley1961detailed} are commonly used for long-term assessment of sustainable energy policies. Shannon's limit of information capacity \cite{shannon1959coding} has formed a foundation for the development of modern communication systems. The Bode's integral on sensitivity limits in feedback control theory \cite{seron1997limitations} is a standard tool for analysis of design trade-offs in modern control systems. At the same time, the question of what are the fundamental limitations of nonlinear harvesting efficiency is still open.

There have been only few studis that addressed the question of the maximal power limit. In \cite{mitcheson2004architectures,stephen2006energy} the maximum power limit was derived for harmonically excited specific designs of velocity-damped and coulomb-damped resonant generators as well as for coulomb-force parametric generator. It was also shown in the literature that for harmonic excitation, an optimized nonlinear harvester has a power output of $\frac{4}{\pi}$ times that of a tuned linear harvester\cite{ramlan2010potential}. More recently, maximal power limits for linear\cite{halvorsen2008energy} and nonlinear\cite{halvorsen2013fundamental} energy harvesters under white noise excitation were explored. All the studies conducted thus far exploring the power limits, have considered either specific harvester designs or particular input excitations (mainly harmonic or white noise random vibration).

In this work we provide an analytical framework for deriving the energy harvesting limits for generic nonlinear harvesters excited by arbitrary external forcing. To illustrate the approach, we build a hierarchy of increasingly more constrained models of nonlinear harvesters, derive the closed form solutions for simplest models and provide general formulations where the closed form solutions do not exist. Inspired by the optimal solutions to the simple model we propose a conceptually design of latch-assisted nonlinear harvesters and show that they are significantly more efficient than the traditional linear and nonlinear harvesters in broadband low-frequency excitation that is common to practical situations.

\section*{Ideal Energy harvesting}
We consider a model of a single-degree-of-freedom ideal energy harvester characterized by the mass $m$ and the displacement $x(t)$ that is subjected to the energy harvesting force $f(t)$ and exogenous excitation force $F(t)$. The dynamic equation of the system is a Newton's second law $m\ddot{x}(t)=F(t)+f(t)$. The fluxes of energy in the system are given  by the expressions $F\dot{x}$, $-f\dot{x}$, and $\frac{m}{2}\dot{x}^2$ representing respectively the external input power to the system, harvested power, and instantaneous kinetic energy of the system.

We start our analysis by considering an idealized harvester with no constraints imposed on either the harvesting force $f(t)$ or the displacement $x(t)$. It is easy to show that overall harvesting rate in this setting is unbounded. Indeed, the trajectory defined by a simple relation $\dot{x}(t)=\kappa F(t)$ that can be realized with the harvesting force $f = m\kappa\dot{F} - F$ results in the harvesting rate of $\kappa F^2$ that can be made arbitrarily large by increasing the mobility constant $kappa$. This trivial observation illustrates that the question of fundamental limits is only well posed for the model that incorporate some technological or physical constraints. This is a general observation that applies to most of the known fundamental limits. For example, Carnot cycle limits the efficiency of cycles with bounded working fluid temperature, and Shannon capacity defines the limits for signals with bounded amplitudes and bandwidth.

To derive first nontrivial limits to the energy harvesting efficiencies we consider the displacement amplitude and energy dissipation constraints that are common to all energy harvesters. For the first constrained model we consider the displacement constraint with the trajectory limited in a symmetric fashion, i.e. $|x(t)|\leq x_{\mathrm{max}}$, where $x_{\mathrm{max}}$ is the displacement limit. In this model we assume there is no natural dissipation of energy in the system, so in the steady state motion, the integral net input of energy into the system equals the harvested energy. Thus, the maximum harvested energy could be evaluated simply by maximizing the following expression:
\begin{eqnarray}\label{Eq:1}
   E_{\mathrm{max}} = \max_{x(t)} \int\mathrm{d}t \, F(t)\dot{x}(t).
\end{eqnarray}
Here the optimization is carried over the set of all ``reachable'' trajectories, that can be realized given the system constraints. As long as the harvesting force $f$ is not subjected to any constraints, this set simply coincides with the set of bounded trajectories defined by $|x(t)|\leq x_{\mathrm{max}}$. The optimal trajectory is then easily found by rewriting the integral in Eq.(\ref{Eq:1})  as $-\int \mathrm{d}t \, \dot{F}(t)x(t)\, $. It is straightforward to check that this expression  is maximized by
\begin{align}\label{Eq:2}
 x_*(t)=-x_{\mathrm{max}} \, \mathrm{sign}\left[\dot{F}(t)\right].
\end{align}
The interpretation of Eq.(\ref{Eq:2}) is straightforward and can be summarized as ``buy low, sell high'' harvesting strategy. The optimal harvester keeps the mass at its lowest position until the force $F$ reaches its local maximum at $t=t_0$ (where the sign of $\dot{F}$ changes) and then activates the force $f$ to move the mass by $2 x_{max}$ upwards as fast as possible. The energy harvested during this transition is equal to $2 F(t_0) x_{max}$. Then, the harvester waits until the force reaches its local minimum and repeats the movement in reversed fashion. The process is repeated at every extremum of the external force field. Notably, the infinitely fast motion of the mass at the extremal points implies that the harvester injects energy in the system during the acceleration, and takes it away at the deceleration stages. However, the net balance of energy injections is positive.

This strategy is remarkably similar to the strategy employed by Carnot cycle machine and can be also derived using similar geometric arguments. In the $F,x$ parametric plane, the overall harvested energy is defined as the integral $\oint F dx$ representing the area of the contour produced by the cycle. For a local realization of the force both the values of the force and the values of displacement are bounded, so the energy is maximized by the contour with rectangular shape. Very similarly the Carnot cycle has a simple rectangular shape in temperature-entropy $T-S$ diagram that can be derived by  recognizing that the overall work given by $\oint T dS$ is the area of the contour that is constrained by the temperature limits.

The net harvested energy in this model can be expressed as $E_{\mathrm{max}} = x_{max} \int |\dot{F}(t)| dt$. For commonly used Gaussian models of the random external forces characterized by the Fourier transform $F_\omega = \int \mathrm{dt}\exp(i\omega t) F(t)$, and corresponding power spectral density $|F_{\omega}|^2$, the quantity $\dot{F}(t)$ is a Gaussian random variable with zero mean and the variance given $\int \frac{d\omega}{2\pi}\omega^2 |F_\omega|^2$. Therefore, the maximal harvesting energy is given by the following simple expression:
\begin{equation}
 E_{\mathrm{max}} =x_\mathrm{max} \frac{2}{\pi} \sqrt{\int \frac{\mathrm{d}\omega}{2\pi}\omega^2 |F_\omega|^2}
\end{equation}
The strategy favours the high frequency harmonics which produce frequent extrema of the external force each coming with the harvesting opportunity. Obviously, in practice harvesting energy at very high harmonics will not work because of the natural energy dissipation in the system. So, in our next model, we consider the limits associated with dissipation.

To make the analysis tractable we define a new model without the displacement constraints (so $x_\mathrm{max} = \infty$), but with additional damping force $F_d = -c_m \dot{x}$. Consequently, the dynamic equation changes to $m\ddot{x}(t)+c_m\dot{x}(t)=F(t)+f(t)$, and $c_m \dot{x}^2(t)$ represents the power dissipated in the mechanical damper. The  harvested energy $-\int \mathrm{d}t \,f(t)\dot{x}(t)$ is then equal to $\int\mathrm{d}t \,[F(t)\dot{x}(t)-c_m \dot{x}^2(t)]$, assuming no accumulation of energy in the system at steady state. This is a simple quadratic function in $\dot{x}$ that is maximized by $\dot{x} = F/2c_m$ thus resulting in the following integral energy expression.
\begin{eqnarray}\label{Eq:3}
   E_{\mathrm{max}} = \max \int \mathrm{d}t\left[F(t)\dot{x}-c_m \dot{x}^2\right]
= \int\frac{F^2(t)}{4 c_m}dt.
\end{eqnarray}
As in the previous models, without any constraints on the harvesting force, the trajectory is achievable with the input harvesting force of the form $f(t)=m\ddot{x}_*(t) - F(t)/2$. Furthermore, in view of Parseval's theorem and the final result in Eq.(\ref{Eq:3}), the maximum energy in frequency domain is equal to $E_{\mathrm{max}} = \int \frac{d\omega}{8\pi c_m} |F_\omega|^2$. This simple frequency-domain representation has an important property that with the optimal and ideal harvester force, energy is harvested from all the frequency components of the excitation force equally proportionate to the power spectrum of the forcing function. This is very advantageous to low-frequency and broadband vibration sources such as wave or walking motion where efficient resonant harvesting is not possible. 

In a similar fashion it is possible to construct more complicated limits that combine multiple constraints. Although most of these models do not admit a closed-form solution, the corresponding optimization problem is computationally simple and can be transformed into a system of ordinary differential equations using the Lagrangian multiplier and slack variable techniques. For example, incorporation of the displacement constraints into a damped harvesting model can be accomplished by solving the following variational problem:
\begin{equation}
 E_{\mathrm{max}} = \max \int dt \left[F \dot{x}- c_m \dot{x}^2 - \mu {\cal E} -\lambda ({\cal I}-\alpha^2)\right].
\end{equation}
Here, the unconstrained optimization is carried over $x(t)$, $f(t)$, the two Lagrangian multiplier functions $\lambda(t)$ and $\mu(t)$ and the so-called slack variable $\alpha(t)$. The function ${\cal E}(x,\dot{x},\ddot{x},t) = m\ddot{x} + c_m \dot{x} - F - f$ represents the equality constraint associated with the equations of motion, while the indicator function ${\cal I}(x) = x_{max}^2-x^2$ that is positive only on admissible domain represents the inequality constraint for the displacement. The incorporation of the slack variable $\alpha$ in quadratic form ensures that the inequality constraint is always satisfied. Other equality and inequality constraints on the displacement, velocity, or harvesting force amplitudes can be naturally incorporated in a similar way. Using the standard Euler-Lagrangian variational approach the problem can be transformed into a system of ordinary differential equations that can be easily solved for arbitrary forcing functions and thus provide universal benchmarks for any practical harvesters. It is worth noting, that the general approach of studying the extremal behavior of the physical systems using variational approach is by no means new. In its modern form it originated in the quantum field theory \cite{novikov1983instanton} but has since been applied in many fields most notably in one of the most difficult nonlinear problem of turbulent dynamics \cite{falkovich1996instantons}.

\section*{Force constraints}
Having a totally arbitrary harvesting force as in an ideal harvester, may not be easily realizable with the current technology at least on small scales. More accurate efficiency limits can be derived on models incorporating additional constraints on the harvesting force $f(t)$. In a more realistic representation of the system the harvesting force $f(t)$ can be decomposed into three parts. First, is an inherent or intentionally introduced restoring force from the potential energy $U(x)$ usually originating from the mechanical strain of a deflected cantilever harvester. Second is the linear harvesting energy force equal to $c_e \dot{x}$ that is typical to most of the traditional conversion mechanisms. Finally, controlled harvesters may also utilize an additional control force  $u(t)$ to enhance the energy harvesting efficiency. In contrast to the ideal harvester the control force can not be used for direct extraction of energy from the system, however it can be used to change the dynamics of the system in a way that increases the overall conversion rate $c_e \dot{x}^2$. More precisely, the overall energy harvested from the system is given by  $\int dt [c_e \dot{x}^2 - w(t)]$, where $w(t)$ represents the power necessary to produce the control force $u(t)$ and the corresponding power $p(t) = u(t)\dot{x}$. The corresponding optimization problem can be written as
\begin{equation}
 E_{\mathrm{max}} = \max \int dt\left[F \dot{x} - c_m \dot{x}^2 - l(t)\right].
\end{equation}
Here the new function $l(t) = w(t) - p(t)$ represents the losses of power during the control process. The specifics of the losses process depend on the details of the system design and can be difficult to analyze in a general setting. However, it is easy to incorporate a number of common natural and technological constraints on the loss rate. First, the second law of thermodynamics implies that the losses are always positive. If the the control system cannot accumulate any energy, this constraint can be represented simply as $l(t) \geq 0$. If energy accumulation is possible, only integral constraint can be enforced: $\int l(t) dt \geq0$. Obviously, if this is the only constraint imposed on the system, the optimal solution would correspond to zero losses $l = 0$ and coincide with previous analysis of an ideal harvester.

More interesting bounds can be obtained by incorporating common technological constraints. The obvious one is the introduction of limits on the force value $u_\mathrm{min}\leq u(t)\leq u_\mathrm{max}$ that can be naturally added via additional slack variables as described above. The two other constraints represent different levels of sophistication of the harvesting control systems. First, is the inability of the control system to harvest the energy. Typically the conversion of mechanical energy to useful electrical one happens only through the electric damping mechanism characterized by the force $c_e \dot{x}$. In this case, the work done to produce the control input is constrained to be positive, so $w(t) \geq 0$ or $l(t)\geq -u \dot{x}$. This setup corresponds to a harvesting system where the control force $u(t)$ can inject the energy into the system but cannot extract it from the system. Even more restrictive constraint would correspond to a situation where the control system cannot inject energy at all, so it is only capable of increasing the natural dissipation rate, thus acting as an effective break. In this case, the power injection can be only negative, so $u(t)\dot{x} \leq 0$.

These two extensions of the problem can be naturally transformed into nonlinear systems of differential equations using the slack variable technique explained above. Numerical analysis of these equation may provide upper bounds on the energy efficiency. Comparison of different bounds would then provide a natural way of valuing the potential benefits of possible control systems used in energy harvesters.

\section*{Latch-assisted energy harvesting}
To illustrate the usefulness of the harvesting efficiency limits we propose a conceptually novel way of nonlinear harvester that is inspired by the behaviour of an ideal harvester with no mechanical damping described by (Eq.(\ref{Eq:2})). This harvester is based on a passive control system and satisfies the strictest constraints introduced in our work. At the same time, we show that in practical settings it can achieve very high efficiencies close to the fundamental limits defined by the weakly constrained system. The harvester is based on a simple extension of a classical linear mass-spring damper system with a simple latch mechanism that can controllably keep the system close to $x = \pm x_\mathrm{max}$ positions mimicking the ideal harvester and to enforce the trajectory expressed by Eq.(\ref{Eq:2}).

More specifically, we use a simple control strategy when the secondary stiff spring representing the latch is activated when the harvester mass reaches its maximum or minimum displacement limit, so that the harvester mass is held at the limit. When the force reaches its extremal value a signal is sent to the latch mechanism to release the mass by detaching the secondary spring. 
Dynamic equation of this system could be rewritten as $m\ddot{x}(t)+(c_m+c_e)\dot{x}(t)+U_0'(x)=F(t)-U_l'(x)\sigma(t)$ where $\sigma(t)$ is the signal for activation or deactivation of the latch system. $U_0(x)$ and $U_l(x)$ are respectively the potential energy of the harvester's linear restoring force and the latch mechanism.

Fig.1 illustrates the concept of maximizing the harvested energy through a latch mechanism as one method to mimic the trajectory in Eq.(\ref{Eq:2}). In this method, almost all the work is done on the system when the system is moving from one end to the other; this energy is then harvested and dissipated when the system is blocked by a latch from moving outside of the extremal points. Whenever the excitation is slow in comparison to the natural period of the harvester, the system translates between the extrema very fast, while the force remains close to its extremal values. The system takes natural advantage of the frequencies and unlike traditional linear harvesters has a higher efficiency at low frequencies.

To compare with the alternative approaches to nonlinear harvesting, we choose one of the most efficient and popular harvesters with the bistable potential of the form $U(x) = \frac{1}{2}k_1(1-r) x^2 -\frac{1}{4}k_3 x^4$. Here, $k_1$ is the linear spring constant, $k_3$ is the nonlinear spring constant, and $r$ is a tuning parameter. Fig.2 compares the three equivalent systems differing only in form of potential and control input in terms of displacement and harvested energy when they are excited by base excitation. Fig.2 (b) shows that the latch-assisted harvester outperforms both linear and bistable system. Notably, in this experiment the high performance of latching harvester is achieved without any tuning of the parameters. At the same time the performance of the bistable harvester requires some tuning.

Fig.3 gives further insight to the origin of high energy harvesting efficiencies where we plot the trajectories in the phase space of the system. Fig.3(a) illustrates phase diagrams of the three harvesters. According to the figure, translation between the two ends occur at the largest speed in the latch-assisted harvester that could be indicative of better energy harvesting. Fig.3(b) is even more illustrative showing the force capable of doing positive work versus displacement. The ideal harvester with no mechanical damping, will have a rectangular force-displacement curve on this diagram. The area enclosed by this curve is a better indicator of the harvested energy. This area is maximized for the ideal harvester, and all other harvesters including linear, bistable and latch-assisted will fall inside this rectangle enclosing a smaller area. Revisiting the Carnot cycle analogy, this diagram for energy harvesters is parallel to the temperature-entropy ($T-s$) diagram for heat engines; the rectangle in Fig.3 (b) pertaining to the ideal harvester with only displacement constraint (no mechanical damping) is analogous to the rectangle (the consecutive adiabatic and isothermal processes) in ($T-s$) diagram pertaining to Carnot cycle. In both force-displacement and temperature-entropy diagrams, the enclosed area represents the useful energy or work that is intended to be maximized.

In order to see how efficient the three systems are with respect to the ideal harvester with no mechanical damping, it is convenient to look at the normalized power contours for different values of excitation frequency and amplitude as shown in Fig. 4. According to the figure, linear harvester works well only near the resonance frequency provided that it does not hit the displacement limits. Bistable system has a wider effective region as compared to the linear one. The latch-assisted system has a much larger effective region in terms of excitation amplitude and frequency where it harvests better than linear and bistable systems. The latch system works best at low frequencies and large amplitudes where it can mimic the ideal harvester best. Low efficiency of the latch system in the lower left of the plot (low frequency and small amplitude) is because the system does not reach the displacement limits to latch, and hence works like a linear system in this region. Remarkably the system achieves almost $80\%$ of the maximal efficiency\footnote{The maximal efficiency here refers to the efficiency of the ideal harvester with no mechanical damping. 80 percent of efficiency reported here along with the efficiencies reported in Fig. 4 are calculated with respect to this maximal efficiency. However, if the efficiency of the latch-assisted harvester is calculated with respect to the actual power limit of a harvester with mechanical damping, the efficiency values will be even higher.} in the most interesting regions. This implies that the benefit of introducing more sophisticated active control systems is small, as the system performance is mainly dominated by the displacement constraints.

Despite the huge amount of research and the common belief that the purposeful inclusion of nonlinearity could increase the power density or bandwidth of the vibratory energy harvesters, it has been shown that the current nonlinear harvesters in particular, the bistable harvesters are sensitive to the type of ambient excitation to which they are subjected, and may or may not be very effective when real ambient vibration sources are used \cite{green2013energy}. To analyze how robust the latch-assisted harvester is, we tested its performance on real experimental data of walking motion at the hip level \cite{kluger2014nonlinear} which is inherently a low-frequency motion. According to Fig. 5, the latch-assisted system harvests energy much better than the other two systems.

In conclusion, we have developed a general analysis framework and model hierarchy for derivation of fundamental limits of nonlinear energy harvesting rate. The framework allows easy incorporation of almost any constraints and arbitrary forcing statistics and represents the maximal harvesting rate as a solution of low order system of ordinary differential equations. Closed-form expressions were derived for two cases of harvesters constrained by mechanical damping and maximal displacement lengths. To illustrate the value of the limits we have proposed a simple concept for nonlinear energy harvesting that mimics the performance of the optimal system using a passive latch mechanism. The proposed mechanism outperforms both linear and bistable harvesters in a wide range of parameters including the most interesting regime of low-frequency large-amplitude excitations where the current harvesters fail to achieve high performance.

\textbf{References}
\bibliography{main}

\begin{addendum}
 \item[Competing Interests] The authors declare that they have no
competing financial interests.
 \item[Correspondence] Correspondence and requests for materials
should be addressed to Ashkan Haji Hosseinloo ~(email: ashkanhh@mit.edu).
\end{addendum}

\begin{figure}
  \includegraphics[width = \linewidth]{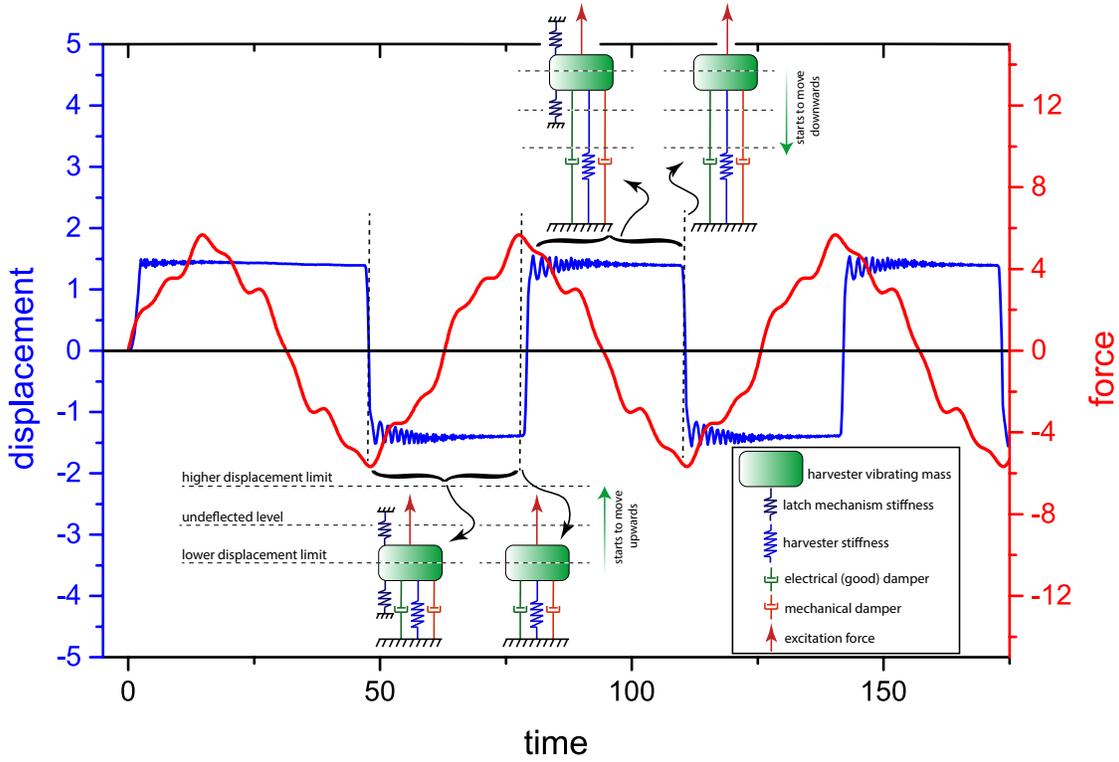}
  \caption{\textbf{Latch-assisted harvester.} Here, an energy harvester with linear mechanical and electrical damping ($\frac{c_m}{2\sqrt{km}}=0.02\: \&\: \frac{c_e}{2\sqrt{km}}=0.1$), and linear stiffness ($k$) is considered. Vibration travel is constrained to 1.5 units i.e. $|x(t)|\leq 1.5$. A much stiffer spring (that is not shown in the figure) is used to guarantee no motion beyond the displacement limits. This could model the mechanical stoppers or the container walls in a real set-up. A direct periodic (multi-frequency) forcing function with dominant frequency of 0.1 is applied. Time and frequencies are nondimensionalized with respect to the undamped natural frequency of the harvester ($\sqrt{k/m}$). Displacement is normalized with respect to a scale length that is set equal to the stable positions of a symmetric double-well potential (a bistable system) that is briefly discussed later in the article.}
  \label{Fig:1}
\end{figure}

\begin{figure}
  \includegraphics[width = \linewidth]{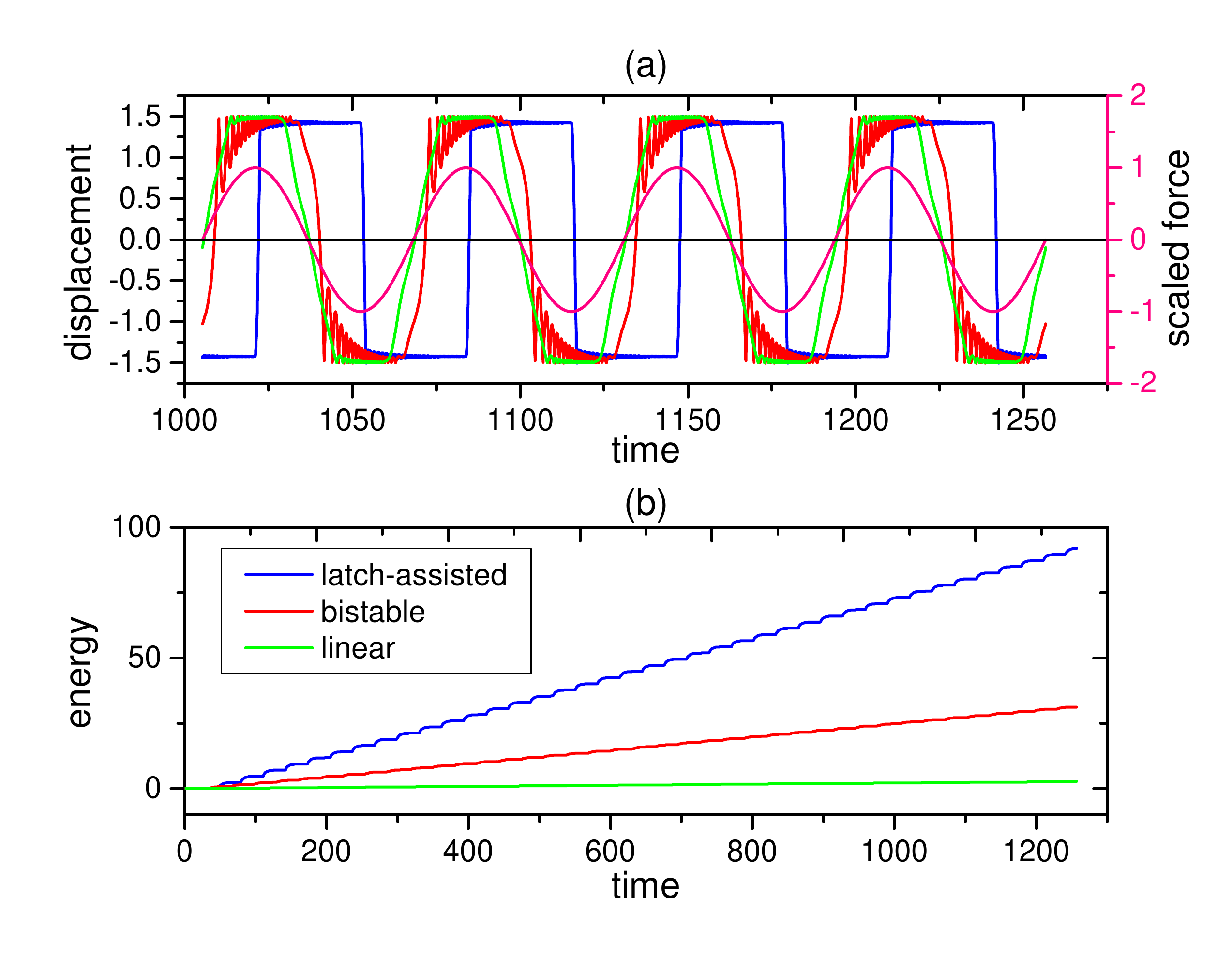}
  \caption{\textbf{Displacement and energy response to harmonic base excitation.} Three linear, bistable and latch-assisted systems are compared when base-excited by a harmonic input. Nondimensional base excitation amplitude and frequency are 200 and 0.1, respectively. The same damping and displacement limit as in Fig. 1 are used. System is simulated for 400$\pi$ units of time i.e. 20 times the excitation period. \textbf{a}, Displacement time history is shown only for four excitation periods. Nondimensional base excitation force is further normalized to unity. \textbf{b}, Nondimensional harvested energy is shown for twenty excitation periods.}
  \label{Fig:2}
\end{figure}

\begin{figure}
  \includegraphics[width = \linewidth]{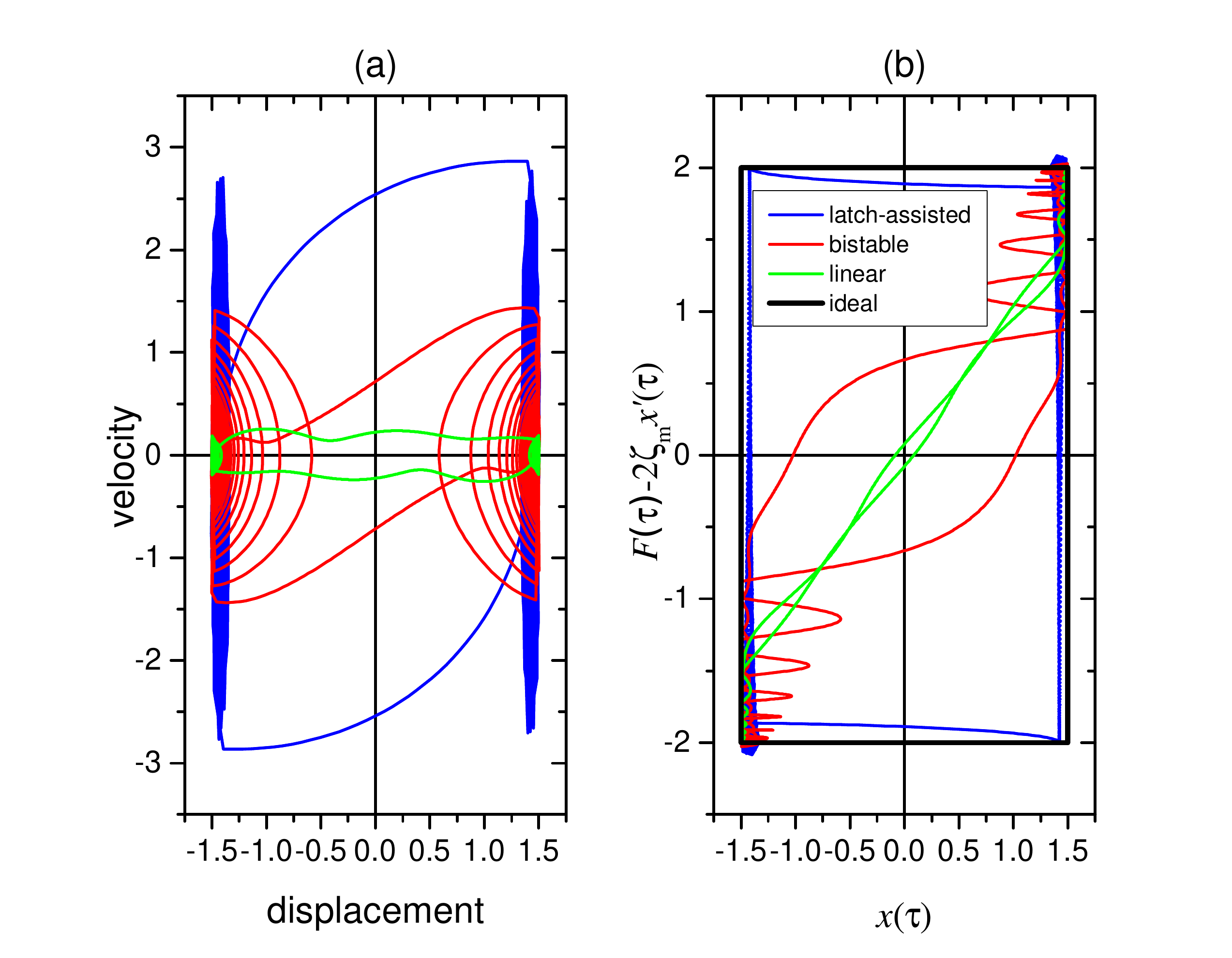}
  \caption{\textbf{Phase and force-displacement diagrams.} \textbf{a}, Depicts the phase diagram for the three linear, bistable, and latch-assisted systems. The same damping and displacement limit as in Fig. 1 are used. \textbf{b}, Depicts the force-displacement curves for the linear, bistable, latch-assisted mechanism, and ideal harvester with no mechanical damping. Here $x$,$F$, and $\tau$ are nondimensional displacement, excitation force, and time. $\zeta_m$ is defined as $\frac{c_m}{2\sqrt{km}}$, and $(.)'$ represents differentiation with respect to the nondimensional time.}
  \label{Fig:3}
\end{figure}

\begin{figure}
  \includegraphics[width = \linewidth]{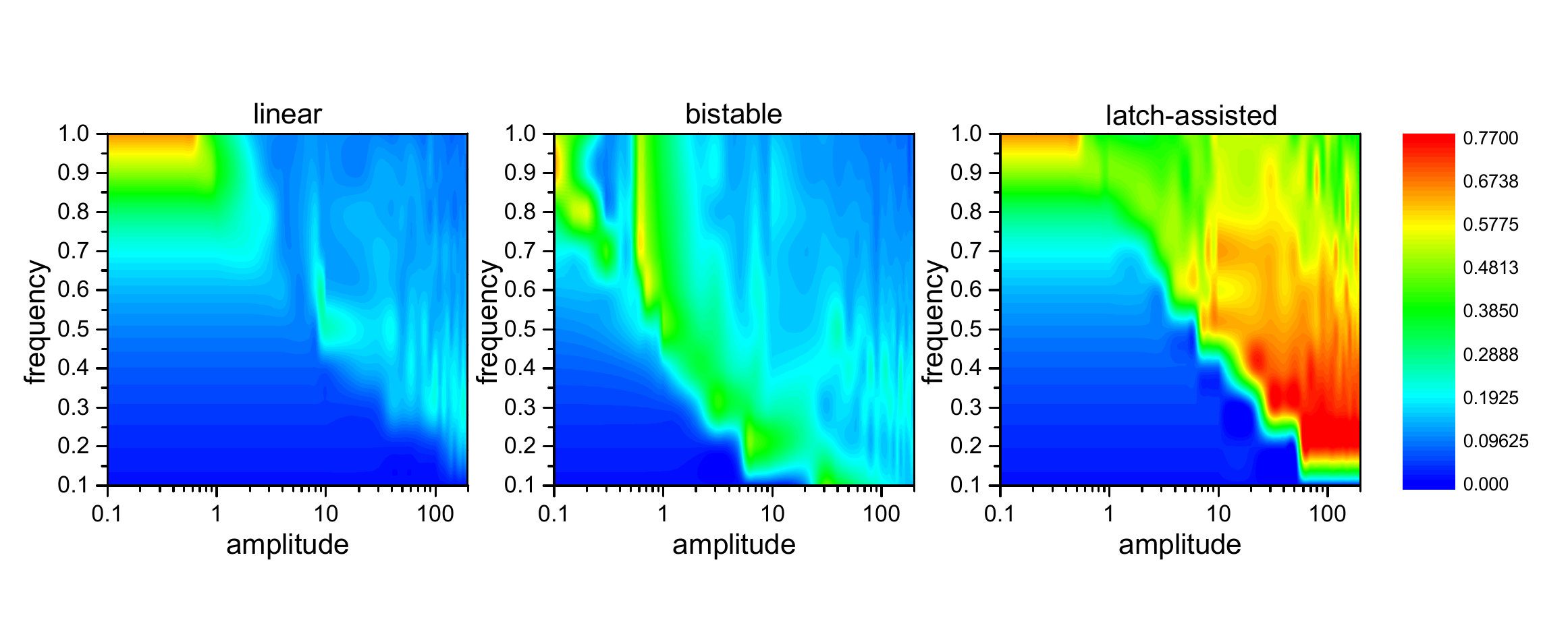}
  \caption{\textbf{Energy harvesting efficiency contours.} Normalized average power of the three systems for a wide range of harmonic base excitation amplitude and frequency is plotted. The average power is calculated by averaging the power over 100 excitation periods at the given excitation frequency. The same damping as in Fig. 1 but different maximum displacement limit of 2.5 units are used. The average power is normalized by the maximum power of ideal harvester with no mechanical damping at the given excitation amplitude and frequency. Using Eqs.({\ref{Eq:1}}) and ({\ref{Eq:2}}), the nondimensional maximum power of the ideal harvester subjected to harmonic base excitation will be equal to $\frac{x_{\mathrm{max}}A\omega^3}{\pi}$, where $A$ and $\omega$ are nondimensional excitation amplitude and frequency, respectively. If the harvester does not reach the displacement limits, actual maximum displacement is used for $x_{\mathrm{max}}$.}
  \label{Fig:4}
\end{figure}
\begin{figure}
  \includegraphics[width = \linewidth]{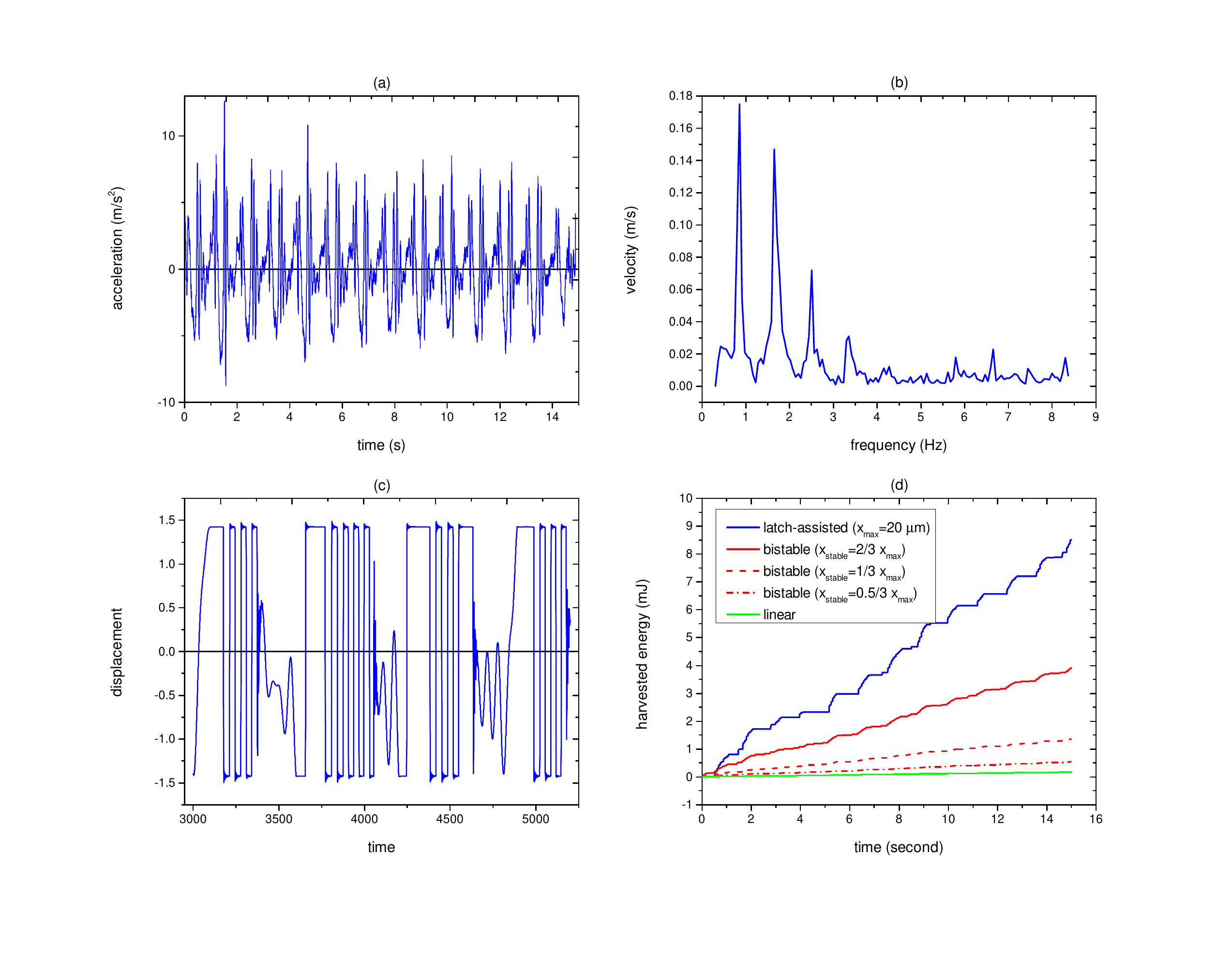}
  \caption{\textbf{Energy harvesting while walking.} \textbf{a}, Experimental acceleration data recorded at the hip while walking \cite{kluger2014nonlinear}. This is an acceleration profile similar to what is experienced by a cellphone one's pocket while walking. \textbf{b}, Velocity spectrum of the walking motion \cite{kluger2014nonlinear}. Since power is proportional to the velocity squared, this could be indicative of power spectrum. \textbf{c}, Partial displacement time history of the nonlinear latch-assisted energy harvester when base-excited by walking motion. The linear natural frequency considered here is 500 $\mathrm{\frac{rad}{s}}$, and the length scale (stable position of bistable system) is set to $\frac{40}{3} \mathrm{\mu m}$. The same damping and normalized displacement limit as in Fig. 1 are used. The time axis is also nondimensionalized. \textbf{d}, Time history of harvested energy for the three systems. A fixed displacement limit of $x_{\mathrm{max}}= 20 \mathrm{\mu m}$ is applied. Three bistable systems with different stable positions of $\frac{2}{3}x_{\mathrm{max}}$, $\frac{1}{3}x_{\mathrm{max}}$, and $\frac{0.5}{3}x_{\mathrm{max}}$ are used. According to the figure, changes in the bistable potential properties could drastically change the bistable harvester efficiency.}
  \label{Fig:5}
\end{figure}

\end{document}